\documentclass[
preprintnumbers]{revtex4}
\usepackage{graphicx}

\begin{document}

\preprint{YITP-03-68}
%\preprint{hep-ph/03?????}

\title{BABAR Resonance and Four-quark Mesons
\footnote{Talk given at the Workshop on 
{\it Progress in Particle Physics}, 
July 22 -- 25,
2003, Yukawa Institute for Theoretical Physics, Kyoto University, 
Kyoto}
} 
%}

\author{K. Terasaki}
\affiliation{Yukawa Institute for Theoretical Physics,
             Kyoto University, Kyoto 606-8502, Japan}

%\date{May 20, 2003}

%\pacs{14.40.Lb, 13.25.Ft}

\begin{abstract}
The new narrow resonance which has been observed at the $B$ factories 
is assigned to the $I_z=0$ component, $\hat F_I^+$, of iso-triplet 
charmed four-quark mesons, and as its consequence, existence of 
additional narrow resonances in different channels is predicted. 
\end{abstract}

\maketitle

%\section{Introduction}
%  \label{sec:Introduction}

Recently the BABAR Collaboration~\cite{BABAR} has observed a narrow 
$D_s^+\pi^0$ resonance with a mass $2317.6 \pm 1.3$ MeV and a width 
$8.8 \pm 1.1$ MeV (a Gaussian fit but the intrinsic width 
$\lesssim 10$ MeV), which has been confirmed by the CLEO\cite{CLEO} 
and the BELLE collaboration\cite{BELLE}, and suggested that it is a 
scalar four-quark meson. 

Four-quark mesons can be classified into four types~\cite{Jaffe}, 
%%%%%%%%%%%%%%%%%%%%%%%%%%%%%%%%%%%%%%%%%%%%%%%%%%%%%%%%%%%%%%%%%%%
$\{qq\bar q\bar q\} = [qq][\bar q\bar q] \oplus (qq)(\bar q\bar q) 
\oplus \{[qq](\bar q\bar q)\pm (qq)[\bar q\bar q]\}$, 
%%%%%%%%%%%%%%%%%%%%%%%%%%%%%%%%%%%%%%%%%%%%%%%%%%%%%%%%%%%%%%%%%%%%
where parentheses and square brackets denote symmetry and
anti-symmetry, respectively. The first two on the right-hand-side 
can have $J^{P(C)}=0^{+(+)}$. Each of them is again classified into 
two classes ({\it heavier} and {\it lighter} ones) because of two 
different ways to produce color singlet states. We discriminate these 
two by putting $\ast$ on the former in accordance with 
Ref.~\cite{Jaffe} in which the four-quark mesons were studied within 
the framework of $q=u,\,d$ and $s$. 

In this report, however, we extend the above framework 
straightforwardly to $q=u,\,d,\,s$ and $c$, and concentrate on the 
$[cq][\bar q\bar q]$ mesons with $q=u,\,d,\,s$. (For more details and 
notations, see Ref.~\cite{Terasaki03}.) The masses of the 
$[cq][\bar q\bar q]$ mesons are now crudely estimated by using a 
simple quark counting with $\Delta m_s = m_s - m_n \simeq 0.1$ GeV, 
($n=u,\, d$), at $\sim 2$ GeV scale and the measured 
$m_{\hat F_I} = 2.32$ GeV as the input data. [We will assign the new 
resonance to $\hat F_I^+$ later.] The estimated mass values are 
listed in Table~I. The masses of the lighter $[cq][\bar q\bar q]$ are 
close to the thresholds of two body decays through strong
interactions. In addition, from the crossing matrices of four-quark
states for the color and the spin~\cite{Jaffe}, it is seen that the 
probability to find the colorless ``$D_s^{+}$'' and ``$\pi^{0}$'' in 
the $\hat F_I^+$ is rather small. Therefore, the rate for the 
$\hat F_I^+\rightarrow D_s^+\pi^0$ can be small as the new resonance, 
although it can decay through iso($I$)-spin conserving interactions. 
Since some of the $[cq][\bar q\bar q]$ mesons are not massive enough 
to decay into two pseudoscalar mesons through $I$-spin conserving 
interactions, their dominant decays may be $I$-spin non-conserving 
ones unless their masses are higher than the expected ones. 
%%%%%%%%%%%%%%%%%%%%%%%%%%%%%%%%%%%%%%%%%%%%%%%%%%%%%%%%%%%%%%%
%\newpage
%%%%%%%%%%%%%%%%%%%%%%%%%%%%%%%%%%%%%%%%%%%%%%%%%%%%%%%%%%%%%%%%
\begin{center}
\begin{quote}
{Table~I. Ideally mixed scalar $[cq][\bar q\bar q]$ mesons 
(with $q=u,\,d,\,s)$, where 
$S$ and $I$ denote the strangeness and the $I$-spin.}
\end{quote}
\vspace{0.5cm}

\begin{tabular}
{|c|c|c|c|c|}
\hline
$\,\, S\,\,$
&$\,\, I=1\,\,$
&$\,\,I={1\over 2}\,\,$
&$\,\, I=0\,\,$
&$\,\,$Mass(GeV)$\,\,$
\\
\hline
$1$
&
\begin{tabular}{c}
$\hat F_I$ \\
$\hat F_I^*$
\end{tabular}
&
&
\begin{tabular}{c}
$\hat F_0$\\
$\hat F_0^*$
\end{tabular}
&
\begin{tabular}{c}
{\hspace{3mm}2.32($\dagger$)}\\
{(3.1)}
\end{tabular}
\\
\hline
0
&
&
\begin{tabular}{c}
{$\hat D$}\\
{$\hat D^*$}\\
{$\hat D^s$}\\
{$\hat D^{s*}$}
\end{tabular}
&
&\begin{tabular}{c}
{2.22}\\
{(3.0)}\\
{2.42}\\
{(3.2)}
\end{tabular}

\\
\hline 
-1
&
&
&\begin{tabular}{c}
$\hat E^0$\\
$\hat E^{0*}$
\end{tabular}
&\begin{tabular}{c}
{2.32}\\
(3.1)
\end{tabular}
\\
\hline 
\end{tabular}\vspace{2mm}\\
\hspace{-30mm}
$(\dagger)$ : Input data
\end{center}
%%%%%%%%%%%%%%%%%%%%%%%%%%%%%%%%%%%%%%%%%%%%%%%%%%%%%%%%%%%%%%%%%%%

Now we study numerically decays of the $[cq][\bar q\bar q]$ mesons by 
assigning the new resonance to $\hat F_I^+$, although there exist 
many proposals to assign it to the other hadron states such as a 
$(DK)$ molecule~\cite{BCL} (or atom~\cite{Szczepaniak}), an 
iso-singlet four-quark meson~\cite{CH}, a $^3P_0\,\,(c\bar s)$ 
state~\cite{CJ}, a chiral partner of $D_s^+$~\cite{chi}, a mixed 
state of a scalar four-quark and a $(c\bar s)$ state~\cite{Browder},
etc. Consider, as an example, a decay, 
%%%%%%%%%%%%%%%%%%%%%%%%%%%%%%%%%%%%%%%%%%%%%%%%%%%%%%%%%%%%%%%%%%%%
$A({\bf p})\rightarrow B({\bf p'})\, +\, \pi({\bf q})$, 
%%%%%%%%%%%%%%%%%%%%%%%%%%%%%%%%%%%%%%%%%%%%%%%%%%%%%%%%%%%%%%%%%%%%
where $A$, $B$ and $\pi$ are a parent scalar, a daughter pseudoscalar 
and a $\pi$ meson, respectively. The rate for the decay is given by 
%%%%%%%%%%%%%%%%%%%%%%%%%%%%%%%%%%%%%%%%%%%%%%%%%%%%%%%%%%%%%%%%%%%%
\begin{equation}
\Gamma(A \rightarrow B\pi)
= {q_c\over 8\pi m_A^2}%\sum_{spin}
|M(A \rightarrow B\pi)|^2,
                                                    \label{eq:rate}
\end{equation} 
%%%%%%%%%%%%%%%%%%%%%%%%%%%%%%%%%%%%%%%%%%%%%%%%%%%%%%%%%%%%%%%%%%%%
where $q_c$ and $M(A \rightarrow B\pi)$ denote the center-of-mass 
momentum of the final mesons and the decay amplitude, respectively. 
To calculate the amplitude, we here use the PCAC hypothesis and a 
hard pion approximation in the infinite momentum frame (IMF), i.e., 
${\bf p}\rightarrow\infty$~\cite{suppl}. 
In this way, the amplitude is given approximately by 
%%%%%%%%%%%%%%%%%%%%%%%%%%%%%%%%%%%%%%%%%%%%%%%%%%%%%%%%%%%%%%%%%%%%
\begin{equation}
M(A \rightarrow B\pi) 
\simeq \Bigl({m_A^2 - m_B^2 \over f_\pi}\Bigr)
                              \langle{B|A_{\bar \pi}|A}\rangle,
                                                     \label{eq:amp}
\end{equation} 
%%%%%%%%%%%%%%%%%%%%%%%%%%%%%%%%%%%%%%%%%%%%%%%%%%%%%%%%%%%%%%%%%%%%
where $A_{\pi}$ is the axial counterpart of the $I$-spin. 
{\it Asymptotic matrix elements} of $A_{\pi}$ can be parameterized by 
using {\it asymptotic flavor symmetry}. (Asymptotic symmetry and its 
fruitful results were reviewed in Ref.~\cite{suppl}.) Although 
asymptotic matrix elements including the four-quark states have been 
parameterized previously~\cite{charm88,charm93,charm99}, we here list 
the related ones, 
%%%%%%%%%%%%%%%%%%%%%%%%%%%%%%%%%%%%%%%%%%%%%%%%%%%%%%%%%%%%%%%%%%%%
\begin{eqnarray}
&&\langle{D_s^+|A_{\pi^-}|\hat F_I^{++}}\rangle 
= \sqrt{2}\langle{D_s^+|A_{\pi^0}|\hat F_I^{+}}\rangle 
= \langle{D_s^+|A_{\pi^+}|\hat F_I^{0}}\rangle
 \nonumber\\
&&=-\langle{D^0|A_{\pi^-}|\hat D^{+}}\rangle            
= 2\langle{D^+|A_{\pi^0}|\hat D^{+}}\rangle 
            \nonumber\\
&&=-2\langle{D^0|A_{\pi^0}|\hat D^{0}}\rangle 
= -\langle{D^+|A_{\pi^+}|\hat D^{0}}\rangle. 
                                                 \label{eq:axial-ch}
\end{eqnarray} 
%%%%%%%%%%%%%%%%%%%%%%%%%%%%%%%%%%%%%%%%%%%%%%%%%%%%%%%%%%%%%%%%%%%% 
Inserting Eq.(\ref{eq:amp}) with Eq.(\ref{eq:axial-ch}) into 
Eq.(\ref{eq:rate}),  we can calculate the approximate rates for the 
allowed two-body decays. Here we equate the calculated width for the 
$\hat F_I^+\rightarrow D_s^+\pi^0$ decay to the measured one, i.e., 
%%%%%%%%%%%%%%%%%%%%%%%%%%%%%%%%%%%%%%%%%%%%%%%%%%%%%%%%%%%%%%%%%%%%
$\Gamma(\hat F_I^+\rightarrow D^+_s\pi^0) \simeq 8.8\quad {\rm MeV}$,  
%%%%%%%%%%%%%%%%%%%%%%%%%%%%%%%%%%%%%%%%%%%%%%%%%%%%%%%%%%%%%%%%%%%%
since we do not find any other decays which can have large rates, and
use it as the input data when we estimate the rates for the other 
decays. The results are listed in Table~II. All the estimated partial 
widths of $\hat F_I$ and $\hat D$ are lying in the region, 4.5 -- 9.0 
MeV, so that they will be observed as narrow resonances in the 
$D_s^+\pi$ and $D\pi$ channels, respectively. 
The $\hat D^s\rightarrow D\eta$ decays are approximately on the 
threshold, i.e., $m_{\hat D^s}\simeq m_D\,+\,m_\eta$, so that it is 
not clear if they are kinematically allowed. 
Besides, the decay is sensitive to the $\eta$-$\eta'$ mixing scheme 
which is still model dependent~\cite{Feldmann}. Therefore, we need 
more precise and reliable values of $m_{\hat D^s}$, $\eta$-$\eta'$ 
mixing parameters and decay constants in the $\eta$-$\eta'$ system to 
obtain a definite result. 

$\hat F_0^+$ cannot decay into $D^+_s\pi^0$ as long as the $I$-spin 
is conserved, so that it will decay dominantly through $I$-spin 
non-conserving interactions. 

$\hat E^0 \sim [cs][\bar u\bar d]$ is an iso-singlet scalar meson 
with $C=1$ and $S=-1$. It cannot decay into $D\bar K$ final states 
unless it is massive enough. If its mass is of almost the same as 
the $\hat F_0^+$, then it cannot decay through strong interactions 
or electromagnetic interactions~\cite{Lipkin,ST} since no ordinary 
meson with $C=1$ and $S=-1$ exists. Therefore, if it can be created, 
it will have a very long life. 
%%%%%%%%%%%%%%%%%%%%%%%%%%%%%%%%%%%%%%%%%%%%%%%%%%%%%%%%%%%%%%%
%\newpage
%%%%%%%%%%%%%%%%%%%%%%%%%%%%%%%%%%%%%%%%%%%%%%%%%%%%%%%%%%%%%%%%
\begin{center}
\begin{quote}
{Table~II. The assumed dominant decays of scalar $[cq][\bar q\bar q]$
mesons and their estimated widths. 
$\Gamma(\hat F_I^+\rightarrow D_s^+\pi^0) = 8.8$ MeV is used as the
input data. The decays into the final states between the angular 
brackets are not allowed kinematically as long as the parent mass 
values in the parentheses are taken.
}

\end{quote}
\vspace{0.5cm}

\begin{tabular}
{|c|c|c|}
\hline
\begin{tabular}{c}
Parent \\
(Mass in GeV)
\end{tabular}
&Final State
&

Width 
(MeV)
\\
\hline
\begin{tabular}{c}
$\hat F_I^{++}(2.32)$ \\
$\hat F_I^+(2.32)$\\
$\hat F_I^0(2.32)$
\end{tabular}
&
\begin{tabular}{c}
$D_s^+\pi^+$\\
$D_s^+\pi^0$\\
$D_s^+\pi^-$
\end{tabular}
& 8.8 
\\
\hline
{$\hat D^+(2.22)$}
&\begin{tabular}{c}
{$D^0\pi^+$}\\
{$D^+\pi^0$}
\end{tabular}
&
\begin{tabular}{c}
{9.0}\\
{4.5}
\end{tabular}
\\
%\hline 
{$\hat D^0(2.22)$}
&\begin{tabular}{c}
{$D^+\pi^-$}\\
{$D^0\pi^0$}
\end{tabular}
&
\begin{tabular}{c}
9.0\\
4.5
\end{tabular}
\\
\hline 
{$\hat D^s(2.42)$}
&$D\eta$ 
&-- 
\\
\hline 
{$\hat F_0^+(2.32)$}
&\begin{tabular}{c}
{$<D_s^+\eta>$}\\
$D_s^+\pi^0$
\end{tabular}
&\begin{tabular}{c}
--\\
($I$-spin non-cons.)
\end{tabular}
\\
\hline 
{$\hat E^0(2.32)$}
&
$<D\bar K>$
&
--
\\
\hline 
\end{tabular}\vspace{2mm}\\
\end{center}
%%%%%%%%%%%%%%%%%%%%%%%%%%%%%%%%%%%%%%%%%%%%%%%%%%%%%%%%%%%%%%%%%%%

In summary we have studied the decays of the scalar 
$[cq][\bar q\bar q]$ mesons into two pseudoscalar mesons by assigning 
the BABAR resonance to $\hat F_I^+$ (in our notation) and assuming 
the $I$-spin conservation. All the allowed decays are not very far 
from the corresponding thresholds so that their rates have been 
expected to saturate approximately their total widths. $\hat F_I$ and 
$\hat D$ could be observed as narrow resonances such as the BABAR
one. To distinguish the present assignment from the other models and 
to confirm it, therefore, it is important to observe these narrow 
resonances. Although we have not studied numerically the 
$\hat D^s\rightarrow D\eta$, we can qualitatively expect that 
$\hat D^s$ will be much narrower than $\hat F_I$ and $\hat D$. 
$\hat E^0$ will decay through weak interactions if it is created 
as long as its mass is below the $\hat E^0\rightarrow D\bar K$
threshold. 

If the existence of four-quark mesons is confirmed, it will be very 
much helpful to understand hadronic weak decays of the $K$ and 
charm mesons. The existence of the heavier class of 
$[qq][\bar q\bar q]$ and $(qq)(\bar q\bar q)$ mesons leads to a 
solution to the long standing puzzle in the charm 
decays~\cite{PDG02}, 
%%%%%%%%%%%%%%%%%%%%%%%%%%%%%%%%%%%%%%%%%%%%%%%%%%%%%%%%%%%%%%
${\Gamma(D^0\rightarrow K^+K^-)/
                  \Gamma(D^0\rightarrow \pi^+\pi^-)}\simeq 3$, 
%%%%%%%%%%%%%%%%%%%%%%%%%%%%%%%%%%%%%%%%%%%%%%%%%%%%%%%%%%%%%%
%%%%%%%%%%%%%%%%%%%%%%%%%%%%%%%%%%%%%%%%%%%%%%%%%%%%%%%%%%%%%%
%\begin{eqnarray}
%&&{\Gamma(D^0\rightarrow K^+K^-)
%                     \over \Gamma(D^0\rightarrow \pi^+\pi^-)}
%\simeq 3,
%\end{eqnarray}
%%%%%%%%%%%%%%%%%%%%%%%%%%%%%%%%%%%%%%%%%%%%%%%%%%%%%%%%%%%%%%
in consistency with the other two-body decays of the charm 
mesons~\cite{charm88,charm93,charm99}. Besides, the lighter 
$(qq)(\bar q\bar q)$  mesons are useful to understand the 
$|\Delta {\bf I}|= 1/2$ rule violating $K\rightarrow \pi\pi$ decays 
in consistency with the $K\rightarrow \pi\pi$ decays satisfying the 
$|\Delta {\bf I}|= 1/2$ rule, the $K_L$-$K_S$ mass difference, the 
$K_L\rightarrow \gamma\gamma$ and the Dalitz decays of 
$K_L$~\cite{Terasaki01}. 

Therefore, it is very much important to confirm the existence of the 
four-quark mesons not only in hadron spectroscopy but also in 
hadronic weak interactions of the $K$ and charm mesons. 

%\section{Conclusion}
%\label{sec:conclusion}

%%%%%%%%%%%%%%%%%%%%%%%%%%%
%\newpage

\section*{Acknowledgments}

The author would like to thank Professor T.~Onogi for providing
information of the BABAR resonance and encouragements. 
He also would like to appreciate  Professor T.~Kunihiro for 
encouragement and kind advises.


\begin{thebibliography}{99}

\bibitem{BABAR} 
BABAR Collaboration, B.~Aubert et al.,  Phys. Rev. Lett. {\bf 90}, 
242001 (2003). 
\bibitem{CLEO} 
CLEO Collaboration, D. Besson et al., hep-ex/0305017
\bibitem{BELLE} 
BELLE Collaboration, K. Abe et al., hep-ex/0307041.
\bibitem{Jaffe} 
R.~L.~Jaffe, Phys Rev. D{\bf 15}, 267 (1977);
{\bf 15}, 281 (1977). 
\bibitem{Terasaki03}
K.~Terasaki, Phys. Rev. D {\bf 68}, 011501(R) (2003). 
\bibitem{BCL} 
T.~Barnes, F.~E.~ Close, and H.~J.~Lipkin, hep-ph/0305025. 
\bibitem{Szczepaniak} 
A.~P.~Szczepaniak, hep-ph/0305060
\bibitem{CH}
H.-Y.~Chen and W.-S.~Hou, hep-ph/0305038.
\bibitem{CJ}
R.~N.~Cahn and J.~D.~Jackson, hep-ph/0305012.
\bibitem{chi}
M.~A.~Nowak, M.~Rho and I.~Zahed, Phys. Rev. D {\bf 48}, 4370 (1993); 
W.~A.~Bardeen and C.~T.~Hill, {\it ibid} {\bf 49}, 409 (1994).  
\bibitem{Browder}
T.~E.~Browder, S.~Pakvasa and A.~A.~Petrov, hep-ph/0307054.
\bibitem{suppl} 
S.~Oneda and K.~Terasaki, Prog. Theor. Phys. Suppl. 
{\bf 82}, 1 (1985). 
\bibitem{charm88} 
K.~Terasaki and S.~Oneda, Phys. Rev. D {\bf 38}, 132 (1988). 
\bibitem{charm93}
K.~Terasaki and S.~Oneda, Phys. Rev. D {\bf 47}, 199 (1993). 
\bibitem{charm99} 
K.~Terasaki, Phys, Rev. D {\bf 59}, 114001 (1999).
\bibitem{Feldmann} 
T. Feldmann, Int. J. Mod. Phys. A {\bf 15}, 159 (2000).
\bibitem{Lipkin}
H.~J.~Lipkin, Phys. Lett. {\bf B70},113 (1977). 
\bibitem{ST}
M.~Suzuki and S.~F.~Tuan, Phys. Lett. {\bf B133}, 125 (1983). 
\bibitem{PDG02}
Particle Data Group, K.~Hagiwara et al, Phys. Rev. D {\bf 66}, 1
(2002). 
\bibitem{Terasaki01} 
K.~Terasaki, Int. J. Mod. Phys. A {\bf 16}, 1605 (2001) and references 
quoted therein.
\end{thebibliography}
\end{document}